\font\elevenrm=cmr10 at 11pt
\font\it=cmti10 at 10pt
\font\bbf=cmbx10 at 11pt

\font\srmm=cmr7 at 7pt

\font\fourteenbf=cmbx10 at 17.28pt

\def\small{\srmm \sevenkmj \baselineskip=8truept}

\def\large{\elevenrm \elevenkmj}
\def\large\bf{\bbf \elevenkmjb}
\def\Large\bf{\fourteenbf\forteenkmjb}

\def\ni{\noindent}
\def\hi{\noindent \hangindent=1.0cm}

\def\pet#1{\vspace{12mm} \centerline{\large \bfseries #1 \vspace{2mm}}}
\def\pen#1{\vspace{4mm} \centerline{\small \bfseries #1 \vspace{-2mm}}}
\def\pea#1{\vspace{2mm} \centerline{\footnotesize #1 \vspace{-2mm}}}
\def\ptd#1{\vspace{5mm} \centerline{\small \it (#1) \vspace{2mm}}}

\def\s#1{\vspace{6mm}{\normalsize \centerline{\bfseries #1}} \vspace{3mm}}

\documentstyle[11pt,twoside,times,bezier,epsf,graphicx,epsfig,rotating,amsmath,amssymb,longtable]{jart-e}
\def\beginabstract{\begin{center}
                   \centerline{\small\bf ABSTRACT}}
\def\beginabst{\begin{center}
               \centerline{\small\bf 요 \ \ \ \ \ 약}}
\def\endabstract{\end{center}}

\newenvironment{abst}{
                 \begin{center}{\small\bf ABSTRACT}\end{center}%
                 \small
                 \vspace{-4mm}%
                 \list{}{\topsep=0cm\parsep=0cm%
                         \rightmargin=1cm\leftmargin=1cm}%
                 \vspace{2mm}
                 \item[]%
                 }{\endlist}%

\begin{document}
\small
\ni {\bf \small J. Astron. Space Sci. 1(1), 1--8 (2013)}
\vspace{-5mm}

\pet{Gamma-ray emission from millisecond pulsars - an Outergap perspective}

\pen{K.S. Cheng}

\pea{Department of Physics, University of Hong Kong, Pokfulam Road, Hong Kong}

\pea{email :hrspksc@hkucc.hku.hk (KSC) }
\ptd{Received ; Accepted }

\begin{abst}
In this review paper we explain the following gamma-ray emission features from the millisecond pulsars. (1)Why is the dipolar field of millisecond pulsars so weak but the magnetic pair creation process may still be able to control the size of the outergap? (2)A sub-GeV pulse component could occur in the vicinity of the radio pulse of millisecond pulsars. (3)Orbital modulated gamma-rays should exist in the black widow systems for large viewing angle.

\end{abst}
\vskip 0.5cm {\it Keywords} :millisecond pulsars, gamma-rays

\vspace{5mm}
\s{1. INTRODUCTION}

Pulsars are known to be rapidly spinning and strongly magnetized neutron stars so they can behave like a unipolar inductor. Without the charge screening young pulsars can easily develop huge potential drop along the open field lines (cf. Manchester \& Taylor 1977; Lyne \& Graham-Smith 1998 for a general review). Such large potential drop can accelerate charged particles to extremely relativistic speed. Since the magnetic field in the open field lines are still sufficiently strong so charged particles are confined to move along the field lines, hence high energy curvature photons are emitted. It has been shown that pulsar magnetosphere is filled with charge separated plasma, whose charge distribution is given by the Goldreich-Julian charge density $\rho =\frac{\vec{\Omega} ~\dot~ \vec{B}(r)}{2\pi c}$, where $\vec{\Omega}$ is the angular velocity vector and $\vec{B}(r)$ is the local magnetic field vector(Goldreich and Julian 1969). When  $\vec{\Omega}$ is perperdicular to $\vec{B}(r)$, the charge density is zero and such region is called null charge surface and charged carriers on two sides of the null charge surface are opposite. With this charge distribution the electric field along the magnetic field is screened out. However Cheng, Ho and Ruderman (1986a) argued that when the current pass through the null charge surface, opposite charged carriers are removed from the vicinity of null charge surface and hence vacuum region will be formed. Charged particles in this vacuum region will be accelerated to extremely relativistic speed so this region is also called "Outergap" accelerator. The question is if the current continue to pass through the null charge region will the vacuum region to grow unlimitedly. The answer is "no" because the energy of the curvature photons is proportional to the size of vacuum region. When the energy of curvature photons is energetically enough, curvature photons and soft photons from the stellar surface can be converted into electron/positron pairs, which can limit the size of the Outergap.  Zhang and Cheng (1997) argued that the soft photons from the surface should be dominated by thermal photons emitted by the hot polar cap, which is heated by the return current from the Outergap. In this particular model, the fractional size of the Outergap ($f_{pp}$) limited by the photon-photon pair creation process is completely determined by the global parameters of pulsars, i.e. rotation period $P$ and the dipolar magnetic field $B$, and it is given by

\begin{equation}
f_{pp}=5.5P^{26/21}B_{12}^{-4/7},
\end{equation}

where $B_{12}$ is the dipolar magnetic field in units of $10^{12}$G and the total gamma-ray luminosity emitted by the Outergap is given by

\begin{equation}
L_{\gamma} \sim f_{gap}^3L_{sd},
\end{equation}

where $f_{gap}$ is the fractional size of the Outergap and $L_{sd}$ is the spin-down power of the pulsar.

It has been shown that many observed features from gamma-ray pulsars can be explained in terms of the Outergap models (e.g. Cheng, Ho and Ruderman 1986b; Romani 1996; Zhang and Cheng 1997; Cheng, Ruderman and Zhang 2000; Takata, Shibata \& Hirotani 2004; Hirotani 2006, 2008; Wang, Takata and Cheng 2010)

\s {2. OUTERGAP CLOSED BY MAGNETIC PAIR CREATION}

Takata, Wang and Cheng (2010) suggest that if the surface multiple field is much stronger than the dipolar field near the stellar surface in particular they can bend those open field lines connecting to the Outergap sideward, then the curvature photons emitted by the return current from the Outergap can become pairs near the surface and their pitch angles can be larger than 90 degrees(cf. Figure 2 of Takata, Wang and Cheng 2010). In this situation, the magnetic pairs can stream back to the outermagnetosphere and restrict the size of the Outergap. The Outergap size restricted by the magnetic pair creation process is given by

\begin{equation}
f_{mp}=0.25K(B_m, s)P_{-1}^{-1/2},
\end{equation}

where $P_{-1}$ is the rotation period in units of $10^{-1}$s,$K(B_m, s)$ depends on the local multiple magnetic field ($B_m$) and the local curvature radius ($s$). By fitting the first Fermi pulsar catalog(Abdo et al. 2010), they find that $K$ is roughly a constant but the value is unity for canonical pulsars and is about 10 for millisecond pulsars respectively. These two values correspond to quadruple field strength $B_m\sim 10^{13}$G for canonical pulsars and $B_m\sim 10^{11}$G for millisecond pulsars respective. They also show that the magnetic pair creation process is more important for pulsars with spin-down power lower than $10^{36}$erg/s. Although the existence of multiple field about 10 times of the dipolar field is possible (e.g. Arons 1983), why could old neutron stars like millisecond pulsars, which is likely older than billion years old, can still maintain a surface multiple field hundred to even thousand times stronger than the dipolar field ? Ruderman (1991) argue that in the core of the neutron star the interpinning between the quantized magnetic flux tubes and the quantized rotation vortex lines is very strong, the spin-down and spin-up of the neutron star can cause the vortex lines to move outward and inward respectively and hence the flux tubes move together with the vortex lines. During the accretion spin-up epoch of the millisecond pulsars the core magnetic field, which cannot decay due to interpinning with the vortex lines, will be squeezed in the rotation axis due to the dragging of vortex lines. Due to the  conservation of magnetic flux the local magnetic field strength will increase but this squeezed magnetic field cannot be stronger than the critical field $B_c \sim 10^{15}$G otherwise the quantized flux tube structure will break down then the magnetic field lines will be reconnected to reduce the local magnetic field strength to below the critical field strength. The structure of this squeezed magnetic field can be illustrated as follow. The vortex lines will drag the north (south) poles of the flux tubes arrive the rotation axis first then the south (north) poles will follow (cf. Fig. 2 of Chen and Ruderman 1993). But these north and south poles cannot be reconnected because underneath these two magnetic poles they are separated by
the vortex lines. Therefore the north will cluster near the rotation axis surrounded by the south poles in the perimenter, a quadruple magnetic field structure is formed. We can estimate the strength of quadruple moment as $Q\sim B_c a^4$, where $B_c$ is the critical field strength and $a$ is the characteristic size of the quadruple moment, which can be estimated by the conservation of vortex lines $N_v=\frac{2\pi r^2 \Omega}{\kappa}$, where $\Omega$ is the angular velocity, $r$ is the radius from the rotation axis and $\kappa$ is the quantized vorticity. We obtain

\begin{equation}
a\sim 3\times 10^4 R_6(P_i/P_{f,-3})^{1/2}cm,
\end{equation}

where $R_6$ is the core radius in units of $10^6$cm, $P_i$ is the initial period before spin-up and $P_{f,-3}$ is the final period after spin-up in units of $10^{-3}$s. It is very important to note that this quadruple moment is located at the boundary between the core and the crust. The magnetic field produced by this quadruple moment on the surface of the neutron star is given by

\begin{equation}
B_m \sim Q(a/l)^4 =10^{11}(B_c/10^15G)(a/3\times 10^4cm)^4(l/3\times 10^5 cm)^{-4},
\end{equation}

where $l$ is the thickness of the stellar crust.

\s {3. SOFT GAMMA-RAY COMPONENT IN MILLISECOND PULSARS}

Cheng, Ruderman and Zhang (2000) have shown that most pairs should be created around the null charge surface. In this Outergap model, the current flow inside the Outergap is dominated by the outflow current from the null charge surface toward the light cylinder whereas the inflow current dominates from the null charge surface toward the star. It has been shown by Hirotani (2005) that the inner boundary of the Outergap is not located at the null charge surface when the outergap current is not zero. Roughly speaking the location of the inner boundary can be estimated as

\begin{equation}
r_{in}\sim r_{null}(1-\frac{J_{gap}}{J_{GJ}}),
\end{equation}

where $r_{in}$ and $r_{null}$ are the distances to the inner boundary and null charge surface respectively, and $J_{gap}$ and $J_{GJ}$ are the gap current and the Goldreich-Julian Current respectively.  The electric field between the null charge surface and the inner boundary can be approximated (Hirotani 2005; Tang et al. 2008) by

\begin{equation}
E_{in}(r)=E(r_{null})\frac{(r/r_{in})^2-1}{(r_{null}/r_{in})^2-1},
\end{equation}

where $E(r_{null})$ is the electric field at the null charge surface. Since the electric field from the null charge surface to the light cylinder is roughly constant whereas the electric field from the null surface to the inner boundary decreases quadratically therefore the characteristic curvature photon energy and the radiation power by the inward current flow are weaker than those of the outflow. The average electric field in the inflow region is $<E_{in}>\sim \int_{r_{in}}^{r_{null}} E(r)dr/r_{null}\sim E(r_{null})/3$, which gives the characteristic curvature photon energy of inflow as

\begin{equation}
E_{\gamma}(in)\sim 3^{-3/4}E_{\gamma}(out),
\end{equation}

where $E_{\gamma}(in)$ and $E_{\gamma}(out)$ are the characteristic curvature photon energies emitted by the inflow and outflow charged particles respectively. We can also estimate the inflow radiation power as

\begin{equation}
L_{\gamma}(in) \sim L_{\gamma}(out) \frac{<E_{in}> r_{null}}{E(r_{null})r_L}\sim (0.15/\rm{tan^2 \alpha})L_{\gamma}(out),
\end{equation}

where $r_L$ is the light cylinder radius and we have used $\frac{r_{null}}{r_L}\approx \frac{4}{9\rm{tan^2 \alpha}}$, where $\alpha$ is the inclination angle. Since the caustic effect of inflow radiation is small, this soft gamma-ray component should occur very near the radio pulse emitted from the polar cap. Unlike in the case of millisecond pulsars  this soft gamma-ray component is very difficult to be observed in canonical pulsars because most of inflow curvature photons will be converted into pairs by the strong magnetic field (cf. Cheng and Zhang 1999; Wang, Takata and Cheng 2013a). Unless the inclination angle and the viewing are both small most inflow curvature photons will be reprocessed into hard X-rays and PSR1509-58 is the representative example (Wang, Takata and Cheng 2013a). The detail fitting of energy dependent light curves of millisecond pulsars can be found in Wang, Takata and Cheng (2013b).

\s {4. ORBITAL MODULATED GAMMA-RAY FROM BLACK WIDOW SYSTEMS}

It is believed that most of pulsar spin-down power eventually will be converted from low frequency electromagnetic dipole radiation into the particle kinetic energy of the pulsar wind. However the exact conversion process from EM wave energy into particle energy is still unclear. In the Crab nebula in a distance $\sim 3\times 10^{17}$ from the pulsar almost all EM wave energy becomes particle kinetic energy (Kennel and Coroniti 1984a, 1984b). On the other hand, by fitting the pulsed TeV data of the Crab pulsar/nebula detected by MAGIC Aharonian et al. (2012) have argued that only a few times of the light cylinder radii from the star a good fraction of spin-down power has already been in the kinetic energy of particles. It is still controversial if the pulsed TeV gamma-rays from the direction of the Crab pulsar are emitted in the magnetosphere or outside the light cylinder. PSR B1259-63/LS 2883 is one of most studied gamma-ray binaries, which is a binary system in which a 48 ms pulsar orbits around a Be star in a high eccentric orbit with a long orbital period of about 3.4 yr. It is special for having asymmetric two-peak profiles in both the X-ray and TeV light curves(Johnston et al. 1994,
1996, 2005; Aharonian et al. 2005, 2009; Chernyakova et al. 2006,
2009; Uchiyama et al. 2009). Recently, an unexpected GeV flare has also been detected by the Fermi gamma-ray observatory several weeks after the last periastron passage (Abdo et al. 2011; Tam et al. 2011). Although X-rays and TeV gamma-rays are generally expected to be emitted from the shock region (Tavani and Arons 1997; Takata and Taam 2009), its aysmmetric light curves are very difficult to be explained. In order to explain its asymmetric two-peak multi-wavelength light curves, Kong et al. (2011; 2012) argue that one of important factors to cause such asymmetric light curve result from the fact that  the particle kinetic energy of the pulsar wind is position dependent. They argue that  the particle kinetic energy of the pulsar wind should gradually increase and can be approximately described by a power law as

\begin{equation}
\sigma(r) = \sigma_{\rm L}(\frac{r}{r_{\rm L}})^{-\beta},
\end{equation}

where $\sigma(r)$ is called the magnetization parameter and is defined as the ratio of the magnetic energy density and the
particle kinetic energy density in the pulsar wind at position $r$, $\sigma_{\rm L}$ is the magnetization parameter at the light cylinder
and $\beta$ is a fitting parameter of order of unity. They estimate the magnetization parameter at the light cylinder as

\begin{equation}
\sigma_{\rm L} = \frac{B^2_{\rm L}/8\pi}{2\dot{N}_{e^\pm} m_{\rm e}
c/r_{\rm L}^2} \sim 4.68 \times 10^4 (\frac{B_{\rm L}}{2.5 \times
10^4 {\rm G}})^2 (\frac{r_{\rm L}}{2.3 \times 10^8 {\rm cm}})^2
(\frac{N_{\rm m}}{10^4})^{-1} (\frac{\dot{N}_{\rm GJ}}{5.26 \times
10^{31} {\rm s}^{-1}})^{-1},
\end{equation}

where $B_{\rm L}$ is the magnetic field at the light cylinder,
$r_{\rm L}$ is the radius of the light cylinder, $\dot{N}_{e^\pm} =
N_{\rm m} \dot{N}_{\rm GJ}$, $N_{\rm m}$ is the $e^\pm$ multiplicity
and $\dot{N}_{\rm GJ} \sim 5.26 \times 10^{31} (B/3 \times 10^{11}
{\rm G}) (P/47.762 {\rm ms})^{-2} {\rm s}^{-1}$ is the
Goldreich-Julian particle flow at the light cylinder. Replacing the characteristic values of PSR B1259-63 by the values of millisecond pulsars
$\sigma_{\rm L}$ is roughly $10^4$. Most black widow systems have an orbital radius of order of $10^{11}$cm and a light cylinder radius of order of $10^7$cm, which imply that a good fraction of pulsar spin-down energy has already been converted into the particle kinetic energy of the pulsar wind in a distance of $10^{11}$cm. The TeV gamma-rays of PSR B1259-63 are produced by inverse Compton scattering. Although the companion stars of black widow systems are white dwarfs, intense optical emission from the companion stars are observed and it is believed to result from the irradiation of pulsars. Therefore it is interesting to ask if high energy photons can be produced via inverse Compton (IC) process and such emission should exihibit orbital modulation. The characteristic energy of IC photons depends on the buld Lorentz factor of the pulsar wind and the characteristic energy of the soft photons from the companion star.

Cheng et al. (2010)
have estimated bulk Lorentz factor of the pulsar wind from millisecond pulsars $\gamma_w$, which is given by

\begin{equation}
\gamma_w = 2\times 10^5 f_{e^{\pm}}^{-1} L_{34}^{1/2},
\end{equation}

where $ L_{34}$ is the pulsar spin-down power in units of $10^{34}$erg/s, $f_{e^{\pm}}= 1 + \frac{m_e \eta_{e^{\pm}}}{m_p}$ and $\eta_{e^{\pm}}=\frac{\dot{N}_{e^{\pm}}}{\dot{N}_p}$ is number ratio between $e^{\pm}$ pairs and protons. For the characteristic values of millisecond pulsars, i.e. $P\sim 3ms$ and $B\sim 3\times 10^8$G, which gives $\gamma_w \sim 3\times 10^4$. Takata, Cheng and Taam (2010; 2012) have suggested that the companion star in the black widow systems will emit optical due to the irradiation of gamma-rays from pulsar. The optical luminosity due to reprocessing the pulsar wind/gamma-rays is given by

\begin{equation}
L_{opt} \sim L_{sd} \frac{\pi R_{ro}^2}{4\pi R_b^2} = 2.5\times 10^{30}L_{34}(10R_{ro}/R_b)^2 {\rm erg/s},
\end{equation}

where $R_{ro}$ and $R_b$ are the radius of the Roche lobe and the binary system respectively. The effective temperature of the companion star is given by

\begin{equation}
T_{eff} = (\frac{L_{opt}}{4\pi R_{ro}^2 \sigma})^{-1/4} = 5\times 10^3 (L_{opt}/3\times10^{31}erg/s)^{1/4}(R_{ro}/10^{10}cm)^{-1/2}K.
\end{equation}

The inverse Compton scattering between the optical photons from the companion star and the pulsar wind can produce high energy photons with the characteristic energy given by

\begin{equation}
E_{IC} \sim \gamma_w^2 (3kT_{eff}) \sim 3GeV (\gamma_w/3\times 10^4)^2(L_{opt}/3\times 10^{31}erg/s)^{1/4}(R_{ro}/10^{10}cm)^{-1/2}.
\end{equation}

The luminosity of the inverse Compton scattering is given by

\begin{equation}
L_{\rm IC}(r) \sim \sigma_{T}n_{ph}(r)r L_{sd}\sim 2\times 10^{-3}(R_b/r)(T_{eff}/5\times 10^3)^3 (R_{ro}/10^{10})^{2}(R_{b}/10^{11})^{-1}L_{\rm sd},
\end{equation}

where $n_{ph}(r)\sim \sigma T_{eff}^3R_{ro}^2/kcr^2$ is the optical photon density at distance $r$ from the companion star. If the line of sight is sufficiently closed to the companion star, i.e. $r\sim R_{ro}$, the maximum inverse Compton luminosity occurs when the companion star is between the observer and the pulsar and it could be about a few percents of the spin-down power and this luminosity falls off
as $1/r$. This component sensitively depends on the viewing angle. Recently the first black widow system, PSR B1957+20 has been shown some evidence of an orbital modulated gamma-ray component at energy >3GeV (Wu et al. 2012)

\s {5. Summary}

We argue that the reason why millisecond pulsars can maintain a surface quadruple field with strength $\sim 10^{11}$G results from interpinning of quantized flux tubes and quantized vortex lines. The magnetic flux tubes are dragged toward the spin-axis during the accretion spin-up phase. Consequently an extremely strong quadruple can be formed in the boundary between the inner crust and the core. This multiple field is very important to ensure the magnetic pair creation process can occur even in millisecond pulsars with dipolar field $\sim 10^8$G. In the Outergap models we predict that a sub-GeV component should exist in the vicinity of radio pulse, which is emitted by the inflow current. But the characteristic energy and luminosity of this component are expected to be lower than the main outflow components. We also predict that an orbital modulated gamma-ray component could be produced by the inverse Compton scattering between the pulsar wind and the optical photons from the companion star.

\vspace{2mm} \ni {\bf ACKNOWLEDGEMENTS:} We thank D. Hui, A. Kong, M. Ruderman, J. Takata, T. Tam and Y. Wang for useful discussion. This work is supported by a GRF grant of Hong Kong Government under 700911P.

\s {REFERENCES}

\hi Abdo, A. A., et al. 2010, ApJS, 187, 460

\hi Abdo, A. A., Ackermann, M., Ajello, M., et al. 2011, ApJ, 736, L11

\hi Aharonian F. et al., 2005, A\&A, 442, 1

\hi Aharonian F. et al., 2009, A\&A, 507, 389

\hi Aharonian, F. A., Bogovalov, S. V., \& Khangulyan, D. 2012, Nature, 482, 507

\hi Cheng, K. S., Ho, C., \& Ruderman, M. A. 1986a, ApJ, 300, 500

\hi Cheng, K. S., Ho, C., \& Ruderman, M. A. 1986b, ApJ, 300, 522

\hi Cheng, K. S., \& Zhang, L. 1999, ApJ, 515, 337

\hi Cheng K.S., Ruderman M \& Zhang L., 2000, ApJ, 537, 964

\hi Cheng, K. S. et al. 2010, ApJ, 723, 1219

\hi Chernyakova M. et al., 2006, MNRAS, 367, 1201

\hi Chernyakova M. et al., 2009, MNRAS, 397, 2123

\hi Goldreich, P., \& Julian, W. H. 1969, ApJ, 157, 869

\hi Hirontani, K. 2005, AdSpR, 35, 1085

\hi Hirotani, K. 2006, ApJ, 652, 1475

\hi Hirotani, K. 2008, ApJ, 688, L25

\hi Johnston S., Manchester R. N., Lyne A. G., Nicastro L., Spyromilio J., 1994,MNRAS, 268, 430

\hi Johnston S., Manchester R. N., Lyne A. G., D…Amico N., Bailes M.,Gaensler B. M., Nicastro L., 1996, MNRAS, 279, 1026

\hi Johnston S., Ball L.,Wang N., Manchester R. N., 2005, MNRAS, 358, 1069

\hi Kennel C. F., \& Coroniti F. V., 1984a, ApJ, 283, 694

\hi Kennel C. F., \& Coroniti F. V., 1984b, ApJ, 283, 710

\hi Kong, S.W., Yu, Y.W., Huang, Y. F., \& Cheng, K. S. 2011, MNRAS, 416, 1067

\hi Kong, S.W., Cheng, K. S. \& Huang, Y. F.,  2012, ApJ, 753, 127

\hi Lyne, A.G. \& Graham-Smith, F. , 1998, Pulsar Astronomy (Cambridge University Press)

\hi Manchester, R. N. \& Taylor, J.H., 1977, Pulsars (Freeman, San Francisco)

\hi Romani, R. W. 1996, ApJ, 470, 469

\hi Takata, J. \& Taam, R., 2009,  ApJ, 702, 100

\hi Takata, J. Cheng, K.S., \& Taam, R., 2010,  ApJ Lett, 723L, 68

\hi Takata, J. Cheng, K.S., \& Taam, R., 2012,  ApJ, 745, 100

\hi Takata, J., Shibata, S., \& Hirotani, K. 2004, MNRAS, 354, 1120

\hi Takata J., Wang Y. \& Cheng K.S., 2010, ApJ, 715, 1318

\hi Tam, P. H. T., Huang, R. H. H., Takata, J., et al. 2011, ApJ, 736, L10

\hi Tavani, M., \& Arons, J. 1997, ApJ, 477, 439

\hi Wang, Y., Takata J. \& Cheng K.S., 2010, ApJ, 720, 178

\hi Wang, Y., Takata J. \& Cheng K.S., 2013a, ApJ, in press (arXiv:1212.2750)

\hi Wang, Y., Takata J. \& Cheng K.S., 2013b, in preparation

\hi Wu, E.M.H. et al. 2012, ApJ, 761, 181

\hi Uchiyama Y., Tanaka T., Takahashi T., Mori K., Nakazawa K., 2009, ApJ, 698, 911

\hi Zhang, L., \& Cheng, K. S., 1997, ApJ, 487, 370

\end{document}